\documentclass[aps,prl,twocolumn,superscriptaddress,nobibnotes]{revtex4}
\usepackage{graphicx}
\usepackage{amsmath}
\usepackage{dcolumn}
\usepackage{bm}
\usepackage{ulem}
\usepackage{times}
\usepackage{amssymb}
\usepackage{epsfig}
\usepackage{xcolor}
\usepackage[T1]{fontenc}
\usepackage{mathtools}   

\begin{document}
\title{Microscopic dynamics of supercooled liquids from first principles}

\author{Liesbeth M.~C.~Janssen}
\altaffiliation{Current address: Institute for Theoretical Physics II,
                Heinrich-Heine University D\"usseldorf, Universit\"atsstra{\ss}e 1, D-40225, Germany}
\email[Electronic mail: ]{ljanssen@thphy.uni-duesseldorf.de}
\author{David R.~Reichman}
\email[Electronic mail: ]{drr2103@columbia.edu}
\affiliation{Department of Chemistry, Columbia University, 3000 Broadway, New York, New York 10027, USA}

\date{\today}

\begin{abstract}
The transition from a liquid to a glass remains one of the most poorly
understood phenomena in condensed matter physics, and still no fully
microscopic theory exists that can describe the dynamics of supercooled liquids
in a quantitative manner over all relevant time scales. Here we present a
theoretical framework that yields near-quantitative accuracy for the
time-dependent correlation functions of a glass-forming system over a broad
density range.  Our approach requires only simple static structural information
as input and is based entirely on first principles. Owing to its ab initio
nature, the framework offers a unique platform to study the relation between
structure and dynamics in glass-forming matter, and paves the way towards a
systematically correctable and ultimately fully quantitative theory of
microscopic glassy dynamics. 
\end{abstract}

\maketitle

Understanding the dynamics of supercooled liquids represents one of the major
challenges in condensed matter science \cite{anderson:95}.  Perhaps the most
striking feature of vitrification is the observed dramatic increase in
viscosity (or relaxation time) upon only a relatively mild change in
thermodynamic control parameters, e.g.\ temperature or density
\cite{debenedetti:01}.  This highly non-linear response of the system is
further accompanied by only subtle changes in the microscopic structure, thus
raising the question as to what physical mechanism underlies the glass
transition \cite{berthier:11,royall:15}. 

Among the various theories of the glass transition proposed in the last few
decades \cite{tarjus:11}, mode-coupling theory (MCT) has acquired a prominent
place in this field of research \cite{gotze:09,das:04,reichman:05}. MCT is the
only strictly first-principles theory of glassy dynamics, and it can accurately
predict many features of the time-dependent dynamics of supercooled liquids
from simple static information alone, including multi-step relaxation patterns
\cite{gotze:09}, stretched exponentials \cite{gotze:09}, and growing dynamical
length scales \cite{biroli:06}.  MCT forms a mean-field framework for broader
theories such as Random First-Order Theory (RFOT) \cite{kirkpatrick:89}, but it
remains a major open question how the relaxation scenarios posited in RFOT can
be captured from first principles and fully microscopically along the same
lines as MCT.  Thus far, efforts to systematically correct the MCT equations,
e.g.\ via higher order field-theoretic loop expansions
\cite{das:86,gotze:87,andreanov:06,miyazaki:05}, have not been explicitly
carried out in a practical setting.  Hence, unlike the situation that arises in
fields such as electronic structure theory, where the correlation energy can be
systematically captured by building upon Hartree-Fock theory
\cite{helgaker:00}, there exists no means of providing a more accurate
description of glassy dynamics by microscopically correcting the lowest-level
mean-field approach. Indeed, a commonly espoused viewpoint is that there exists
a ''no-theory region" between MCT and the true glassy regime \cite{langer:14}.
In this work we show that such a theory can in fact be formulated and
successfully carried out, and we present comparisons with dynamics in a
realistic model of a glassy hard-sphere suspension to quantify the systematic
improvement achieved with our approach.

The key quantity in our discussion is the two-point density correlation
function $F(k,t) = N^{-1} \langle \rho_{\mathbf{-k}}(0) \rho_{\mathbf{k}}(t)
\rangle$, which probes correlated particle motion over time $t$ and at inverse
length scale $k$. Here $\rho_{\mathbf{k}}(t)$ represents a density mode at time
$t$ and wavevector $\mathbf{k}$, $k=|\mathbf{k}|$, $N$ is the total number of
particles, and the brackets denote a canonical ensemble average.  The exact
time evolution of $F(k,t)$ is governed by an integro-differential equation with
memory function $M(k,t)$, the latter of which accounts for time-dependent
damping effects \cite{gotze:09}. More precisely, $M(k,t)$ is related to the
autocorrelation function of a fluctuating force, which captures the effect of
all degrees of freedom orthogonal to the density field \cite{berthier:11}.  It
can be shown that this fluctuating force is dominated by \textit{pairs} of
density modes $\rho_{\mathbf{q}} \rho_{\mathbf{k-q}}$, allowing the memory
function to be expressed in terms of \textit{four}-point density correlation
functions $\langle \rho_{\mathbf{-q}}(0) \rho_{\mathbf{q-k}}(0)
\rho_{\mathbf{q'}}(t) \rho_{\mathbf{k-q'}}(t) \rangle$ \cite{gotze:09}
\footnote{ The time dependence of this four-point correlation function occurs
in a subspace \textit{orthogonal} to the single density mode. However, because
the pair-densities can be made orthogonal to the linear density
\cite{schofield:92}, it is sensible in further approximations to treat the
evolution with unprojected dynamics.}. Within the traditional MCT framework, the
four-point correlation functions are subsequently approximated as products of
two two-point correlation functions, thereby rendering a closed, non-linear
equation of motion for $F(k,t)$ \cite{gotze:09, reichman:05}.  The microscopic
structure of the system enters through the static structure factor $S(k) \equiv
F(k,0)$ \cite{hansen:06}, after Gaussian and convolution approximations are
made for the statics.

In the present work, we seek to avoid the uncontrolled `Gaussian' factorization
of the dynamic four-point correlations, instead retaining these correlations in
the expression for $F(k,t)$ and developing a new exact equation of motion for
the memory term. As shown explicitly in the Supplementary Information (SI) 
\footnote{See Supplemental Material [url], which includes Refs.\ \cite{kob:97,franosch:97,fuchs:91}.}, the
dynamics of these four-point correlation functions is governed by a new memory
kernel that (after projection onto triplet-density modes) contains
\textit{six}-point correlations, i.e.\ correlation functions of the form
$\langle \rho_{\mathbf{k_1}}(0) \rho_{\mathbf{k_2}}(0) \rho_{\mathbf{k_3}}(0)
\rho_{\mathbf{k'_1}}(t) \rho_{\mathbf{k'_2}}(t) \rho_{\mathbf{k'_3}}(t)
\rangle$. These are controlled by eight-point correlations, and so on, thus
allowing one to delay the uncontrolled factorization approximation to a later
stage.  Our approach builds upon the important work of Szamel, who first
introduced such a hierarchical scheme of microscopic kinetic equations
\cite{szamel:03}. Several recent studies have demonstrated that this approach
indeed holds great potential as a microscopic theory of glassy dynamics.  In
particular, it was shown that higher-order correlations can systematically
improve the predicted MCT critical point \cite{szamel:03, wu:05}.  Moreover,
two schematic ($k$-independent) proof-of-principle studies predicted a wealth
of novel relaxation patterns for continuous, discontinuous, and avoided glass
transitions \cite{mayer:06,janssen:14}. For example, the theory may describe
both Arrhenius and super-Arrhenius growth of the relaxation time
\cite{janssen:14}, thus providing potentially the first rigorous
first-principles rationale for the concept of fragility
\cite{angell:95,debenedetti:01}. 

Here we report on the first fully microscopic, high-order calculations to
describe the explicit time- and $k$-dependent dynamics of a realistic glassy
system.  We focus on a system composed of quasi-hard spheres, and compare our
theoretical predictions with the exact dynamics obtained from computer
simulations \cite{weysser:10}.  By using only the simplest measure of static
correlations, i.e.\ $S(k)$, as sole input to our theory, we already achieve
high quantitative accuracy for $F(k,t)$, for all wavevectors, in the regime
from low to moderate supercooling.  In view of its first-principles nature, the
theory can shed important new light on the fundamental connection between
structure and dynamics in glass-forming matter, and paves the way towards an
ultimately fully quantitative description of glassy dynamics.


The first main result of this work is the formulation of a new set of
microscopic equations of motion for the dynamic density correlation functions
of a glass-forming system.  The full derivation is described in the SI.  We
consider the dynamics of the normalized $2n$-point density correlation
functions $\Phi_n(k_1,\hdots,k_n, t)$, which probe particle correlations over
$n$ distinct $k$-values,
\begin{equation}
\Phi_n(k_1,\hdots,k_n, t) = \frac{\langle \rho_{\mathbf{-k_1}}(0) \hdots \rho_{-\mathbf{k_n}}(0)
                                  \rho_{\mathbf{k_1}}(t) \hdots \rho_{\mathbf{k_n}}(t) \rangle}
                                 {\langle \rho_{\mathbf{-k_1}}(0) \hdots \rho_{-\mathbf{k_n}}(0)  
                                  \rho_{\mathbf{k_1}}(0) \hdots \rho_{\mathbf{k_n}}(0) \rangle}.
\end{equation} 
Within our framework, these correlators obey the general equations of
motion
\begin{gather} 
\ddot{\Phi}_n(k_1,\hdots,k_n,t) + \nu\dot{\Phi}_n(k_1,\hdots,k_n,t) 
\nonumber \\ 
+ \Omega^2_n(k_1,\hdots,k_n) \Phi_n(k_1,\hdots,k_n,t) 
\nonumber \\ 
+\int_0^t M_n(k_1,\hdots,k_n,\tau) \dot{\Phi}_n(k_1,\hdots,k_n,t-\tau) d\tau  = 0, \label{eq:GMCTeqPhi_n} 
\end{gather} 
where the dots denote time derivatives, $\nu$ represents a
coefficient that accounts for the short-time dynamics, and the bare frequencies
are given by 
\begin{equation} 
\label{eq:Omega2n}
\Omega^2_n(k_1,\hdots,k_n) = \frac{k_{\rm{B}}T}{m} \left[ \frac{k_1^2}{S(k_1)} + \hdots + \frac{k_n^2}{S(k_n)} \right]. 
\end{equation}
Here $k_{\rm{B}}$ is the Boltzmann constant, $T$ is the temperature, and $m$ is
the particle mass.  For the memory functions we have
\begin{eqnarray}
\label{eq:Mn} 
M_n(k_1,\hdots,k_n,t) = \frac{\rho k_{\rm{B}}T}{16m\pi^3} \sum_{i=1}^n \frac{\Omega^2_1(k_i)}{\Omega^2_n(k_1,\hdots,k_n)}
                  \nonumber \\
                   \times \int d\mathbf{q} |\tilde{V}_{\mathbf{q,k_i-q}}|^2 
                   S(q) S(|\mathbf{k_i-q}|)
                   \hphantom{XXXX}
                  \nonumber \\
                  \times \Phi_{n+1}(q,|\mathbf{k_1-q}\delta_{i,1}|,\hdots,|\mathbf{k_n-q}\delta_{i,n}|,t), \nonumber \\
\end{eqnarray} 
where $\rho$ is the total density, $\delta_{i,j}$ is the Kronecker delta
function, and $\tilde{V}_{\mathbf{q,k_i-q}}$ are static vertices that represent
wavevector-dependent coupling strengths for the higher-order correlations.
These vertices depend on the microscopic structure of the system contained in
$S(k)$, which serves as input to the theory \cite{reichman:05}. Note that the
memory function at level $n$ contains $n$ different terms, each of which
contains a 2($n+1$)-point correlator that measures correlations over a distinct
set of wavenumbers $\{q,k_1,\hdots,|\mathbf{k_i-q}|,k_{i+1},\hdots,k_n\}$.  In
arriving at the above equations, we have retained all diagonal dynamic
$2n$-point correlation functions, and employed Gaussian and convolution
approximations for the static correlators (see SI). These remaining
(uncontrolled) assumptions are essentially the same as at the standard-MCT
level; the crucial difference between our work and standard MCT is that we do
not factorize the diagonal dynamic multi-point correlators. We expect the
inclusion of \textit{off-diagonal} dynamic correlators to lead to a
\textit{matrix} form of the hierarchy that is likely to be far more demanding
computationally, but identical in overall structure compared to the present set
of equations.  We may close this hierarchy of coupled equations at an arbitrary
level $n=N$ by applying a `mean-field' (MF) factorization of the form
$\Phi_N(t) \sim \Phi_{N-1}(t) \Phi_1(t)$ [the simplest example of which is the
standard-MCT closure $\Phi_2(t) \sim \Phi_1(t)\Phi_1(t)$] or by simply setting
$\Phi_N(t) = 0$.  The latter implies that $\Phi_{N-1}(t) \sim
\exp(-\Omega^2_{N-1}t)$, hence we refer to this truncation as an exponential
closure.  Finally, we emphasize that our theory is free of any fitting or
rescaling parameters, thus providing dynamic predictions from purely static
information alone.

\begin{figure}
\epsfig{file=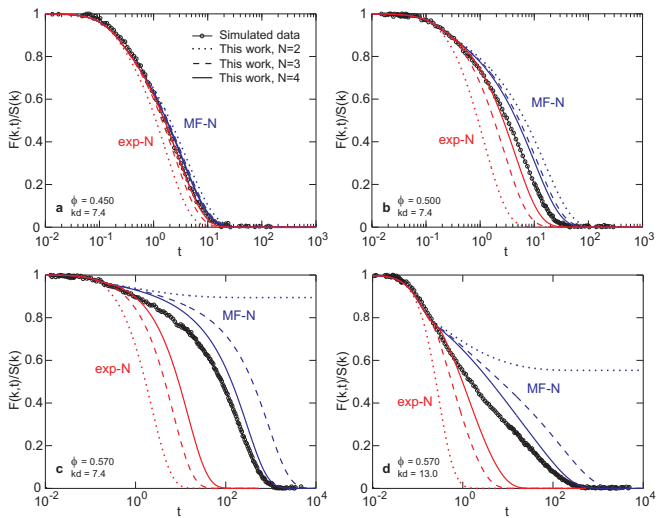,width=0.48\textwidth}
\caption{\label{fig:Fkt} 
Two-point density correlation functions $\Phi_1(k,t)$ for quasi-hard spheres
at different volume fractions and wavevectors:
(\textbf{a}), $\phi=0.450$, $kd=7.4$,
(\textbf{b}), $\phi=0.500$, $kd=7.4$,
(\textbf{c}), $\phi=0.570$, $kd=7.4$,
(\textbf{d}), $\phi=0.570$, $kd=13.0$.
The theoretical results have been obtained using either exponential (red
lines) or MF (blue lines) closures, with closure levels $N=2, 3,$ and 4.  The
$N=2$ MF closure is equivalent to standard MCT. All legends are as in panel
(\textbf{a}). The simulated data (black circles) are taken from Ref.\ \cite{weysser:10}.  
}
\end{figure}

In Fig.\ \ref{fig:Fkt}, we show our numerical results for the density
correlator $\Phi_1(k,t) = F(k,t)/S(k)$ for the quasi-hard sphere system of
Ref.\ \cite{weysser:10}.  While this system is slightly polydisperse to prevent
crystallization, we have used the averaged (effectively monodisperse)
structural properties as input to our calculations, as was also employed by the
authors of Ref.\ \cite{weysser:10}.  As a stringent test of our theory, we
compare the theoretically predicted dynamics with the numerically simulated
$\Phi_1(k,t)$ for different wavevectors and different volume fractions $\phi =
\rho \pi d^3/6$, where $d$ is the average particle diameter. Note that, while
our theory also readily gives access to the higher-order correlators, the
simulated data are limited to $\Phi_1(k,t)$.  The wavevectors we consider
correspond to the main ($kd=7.4$) and second ($kd \approx 13.0$) peak of the
static structure factor.  We have solved the time-dependent coupled equations
numerically employing both MF and exponential closures for truncation levels
$N=2,3,4$.  Computational details can be found in the SI.

Before discussing Fig.\ \ref{fig:Fkt} in detail, we first point out an
important general observation with regard to the convergence of our results.
For all volume fractions and wavevectors considered, we find that the MF
closures provide an \textit{upper} bound to the dynamics, i.e.\ the $N=2$
(standard-MCT) closure predicts the slowest relaxation, while the exponential
closures give a \textit{lower} bound to the dynamics. This is expected because
any mean-field theory will generally underestimate the effects of
ergodicity-restoring fluctuations, consequently overestimating the dynamic
slowing down.  Conversely, an exponential closure rigorously ignores
higher-order memory effects that may drive the system into the glassy state
[since $\Phi_N(t) \sim M_{N+1}(t) = 0$], resulting in relaxation patterns that
are too fast.  With increasing closure level $N$, however, the two types of
closures systematically converge towards each other over an increasingly large
time domain.  This uniform convergence pattern can be rigorously demonstrated
within the confines of a schematic model \cite{mayer:06}, and \textit{provides
a unique and clearly achievable notion of convergence even though no small
parameter exists in our approach}.

Let us now focus on the time-dependent relaxation behavior of the quasi-hard
sphere system for specific values of $\phi$.  For the lowest volume fraction
considered, $\phi=0.450$, the system is still strongly ergodic and
$\Phi_1(k,t)$ exhibits relatively simple relaxation [Fig.\ \ref{fig:Fkt}(a)].
Here our $N=4$ data are almost perfectly converged upon the simulated
$\Phi_1(k,t)$ curve over the complete time domain.  Note that the standard-MCT
result is also reasonably accurate, but slightly overestimates the molecular
relaxation time. These findings hold for all wavevectors considered.  Thus, in
the ''normal liquid" regime, the predictions of our higher-order theory are
virtually exact with respect to computer simulations.

For a higher volume fraction, $\phi=0.500$, the relaxation time is seen to
increase and the system enters the supercooled regime [Fig.\ \ref{fig:Fkt}(b)].
Again we find nearly perfect quantitative agreement between our $N=4$ results
and the simulated density correlator at $kd=7.4$, indicating that our
higher-order microscopic approach accurately captures the onset of glassy
dynamics at the correct absolute value of $\phi$.  This is to be contrasted
with standard MCT ($N=2$), which clearly overestimates the glassiness of the
system and can only reproduce the simulated results for a \textit{rescaled}
volume fraction $\phi^{\rm{MCT}} < 0.500$ \cite{weysser:10}.  Since the
higher-order theory presented here does not require such rescaling of the
control parameter, the inclusion of higher-order dynamic correlations thus
appears to capture all relevant relaxation mechanisms in the weakly supercooled
regime.

We now turn our attention to the highest volume fraction, $\phi=0.570$, where
the system is significantly supercooled. Standard MCT predicts a diverging
relaxation time above $\phi^{\rm{MCT}}_c = 0.566$ \cite{weysser:10}, as seen in
our $N=2$ results in Figs.\ \ref{fig:Fkt}(c)-(d).  The simulated $\Phi_1(k,t)$
curves, however, show complete relaxation to zero, indicating that the system
is still ergodic at this density. In agreement with these simulations, our
$N=3$ and $N=4$ results also predict full relaxation, thus restoring ergodicity
and rounding off the spurious MCT transition.  We find good agreement with the
simulated data for both wavevectors, but our higher-order theoretical results
appear to converge toward a relaxation time that is slightly too fast compared
to the simulations.  The predicted curves for $kd=13.0$ also develop a small
shoulder at intermediate times that is absent in the simulated data.  This
slightly poorer agreement for $kd > 7.4$ was also noted by Weysser \textit{et
al.}\ \cite{weysser:10} for the standard-MCT case, even after rescaling of the
data.  The origin of the discrepancy may be related to the fact that the use of
a density-mode basis neglects some short-time correlations, which are expected
to be more pronounced away from the first peak of $S(k)$.  We attribute further
discrepancies to the remaining approximations in our theory, i.e.\ the neglect
of off-diagonal dynamic correlation functions $\langle
\rho_{\mathbf{q}}(0)\hdots \rho_{\mathbf{q'}}(t) \rangle$ with $\mathbf{q} \neq
\mathbf{q'}$, and the factorization of the static multi-point correlations.
Indeed, one intriguing possibility is that the weaker agreement with simulation
at the highest densities indicates that multi-point \textit{static}
correlations, such as the much studied point-to-set correlations
\cite{berthier:11}, begin to extend a substantial influence on the relaxation.
In this regard, our approach also has utility in delineating the role of static
structural properties beyond $S(k)$ on the dynamics.  Finally, Weysser
\textit{et al.}\ have found that the standard-MCT predictions for this system
can be improved, in particular for small $k$-values, by accounting for
polydispersity effects \cite{weysser:10}. Our theoretical results may thus also
be further improved by careful treatment of the system's polydispersity.

Despite the quantitative differences found in Fig.\ \ref{fig:Fkt}(c)--(d), it
is evident that our higher-order framework in its current form captures a
significant portion of the activated dynamics beyond the standard-MCT scenario,
at least up to moderate volume fractions. We emphasize that our theoretical
predictions are free from any fitting or rescaling procedures, and that regular
MCT would not be able to achieve a similar degree of accuracy, \textit{even
after introducing a rescaling parameter.} The good agreement we find between
our work and simulations is highly non-trivial, considering that the only input
to our theory is $S(k)$, the simplest measure of the microscopic structure of
the system.  Our results thus support the view that activated dynamics--at
least in the weak to moderate regime of supercooling--can be described via a
strictly \textit{dynamic} framework, in the form of deeper memory effects that
occur over increasingly many wavelengths. 

\begin{figure}
    \epsfig{file=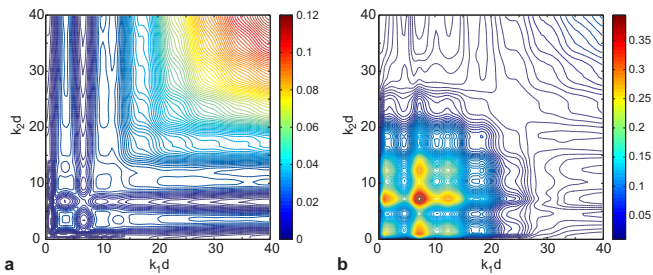,width=0.48\textwidth}
  \caption{\label{fig:F2F1F1} 
  Root-mean-square deviations between the four-point density correlator
  $\Phi_2(k_1,k_2,t)$, as obtained from our high-order calculations
  under an $N=4$ MF closure, and the standard-MCT approximation
  $\Phi^{\rm{MCT}}_1(k_1,t) \Phi^{\rm{MCT}}_1(k_2,t)$, as a function of
  wavevectors $k_1$ and $k_2$ at (\textbf{a}) $\phi=0.450$ and
  (\textbf{b}) $\phi=0.570$.
 }
\end{figure}

As a final part of our analysis, let us elaborate on how our theory improves
upon the standard-MCT framework in terms of the memory function.  Recall that
standard MCT approximates the lowest-order memory contribution
$\Phi_2(q,|\mathbf{k-q}|,t)$ as $\Phi^{\rm{MCT}}_2(q,|\mathbf{k-q}|,t) \sim
\Phi^{\rm{MCT}}_1(q,t)\Phi^{\rm{MCT}}_1(|\mathbf{k-q}|,t)$ \cite{reichman:05}.
Conversely, our framework retains the four-point correlations explicitly and
treats their dynamics rigorously through a hierarchy of equations.
Understanding how the $\Phi_2(q,|\mathbf{k-q}|,t)$ that emerges from our
approach differs from $\Phi^{\rm{MCT}}_2(q,|\mathbf{k-q}|,t)$ thus provides
insight into the improvements afforded over standard MCT.  In Fig.\
\ref{fig:F2F1F1}, we compare our best estimate ($N=4$) for the dynamics of the
four-point correlation function $\Phi_2(k_1,k_2,t)$ with the standard-MCT
approximation $\Phi^{\rm{MCT}}_1(k_1,t) \Phi^{\rm{MCT}}_1(k_2,t)$.  We plot the
root-mean-square deviation (RMSD) $|\Phi_2(k_1,k_2,t)-\Phi^{\rm{MCT}}_1(k_1,t)
\Phi^{\rm{MCT}}_1(k_2,t)|/\sqrt{N_t}$, where $N_t$ is the total number of time
points, for volume fractions $\phi=0.450$ and $\phi=0.570$.  It may be seen
that for both volume fractions the RMSD is modulated by the structure of
$S(k)$.  In the low-density regime [$\phi=0.450$, Fig.\ \ref{fig:F2F1F1}(a)],
the largest deviations are found at \textit{large} wavevectors.  These
differences arise from the interplay between the bare frequency term
$\Omega_2(k_1,k_2)$ and the inertial term $\ddot{\Phi}_2(k_1,k_2,t)$, which
make the four-point correlation functions $\Phi_2(k_1,k_2,t)$ oscillate and
become negative at short times.  Such negative contributions are mostly absent
in standard MCT, since the diagonal MCT terms $\Phi_2(k,k,t) \sim
|\Phi_1(k,t)|^2$ are always positive. Hence, standard MCT tends to overestimate
the memory term $M_1(k,t)$ and consequently overestimates the relaxation time.
These oscillatory patterns only arise in the underdamped limit, however, and
neglecting the inertial term (i.e.\ overdamped dynamics) dramatically reduces
the RMSD in the large-wavevector regime.

In the deeply supercooled region [$\phi=0.570$, Fig.\ \ref{fig:F2F1F1}(b)], the
overall RMSD increases, and the largest deviation is now observed at
\textit{smaller} wavevectors, most notably near the main peak of $S(k)$ at
$k_1d \approx k_2d \approx 7.4$. The RMSD also exhibits local maxima at
wavevectors of $kd \approx 13.0$, corresponding to the second peak of the
structure factor, and $kd \approx 1.0$. Thus, the ergodicity-restoring
processes that round off the standard-MCT transition are mainly contained in
the lower-$k$ behavior of the memory function. It is well established that the
standard-MCT framework is generally the least accurate in the hydrodynamic
$k\rightarrow 0$ limit, resulting in e.g.\ an underestimation of
Stokes-Einstein violation in the deeply supercooled regime \cite{berthier:04}.
Since the higher-order corrections at large volume fractions occur mostly at
small $k$-values, one may expect that our theory might improve the predicted
breakdown of the Stokes-Einstein relation and other related quantities.  This
possibility will be tested in future work.

In conclusion, we have presented a novel first-principles theory to describe
the microscopic dynamics of glassy materials with near-quantitative accuracy in
the low to moderately supercooled regime. The framework requires only static
structure as input and has no free parameters. Our results demonstrate that a
quantitative account of dynamical correlations may be generated from only
two-body statics--even in a regime where many-body statics and dynamical
heterogeneity are markedly growing--, thus shedding light on the connection
between structure and dynamics in supercooled liquids.  Our framework may also
eventually be used to provide predictive insights into regimes of glassy
behavior inaccessible to molecular-dynamics simulations. Not only does our
method give straightforward access to e.g.\ long times and higher-order
correlation functions, it also suggests that by incorporating increasingly many
(static and dynamic) correlations, one may achieve systematically improvable
and ultimately fully quantitative accuracy in the deeply supercooled regime.

As a final note, it should be remarked that the hierarchy presented here
resembles the Martin-Schwinger hierarchy \cite{martin:59,negele:98} in field
theory, where low-order Green's functions are hierarchically connected to
higher-order ones through an infinite set of coupled equations. In the context
of the present work, we point out that there has been substantial recent
progress in the formulation of field-theoretic approaches for the glass
transition that allow for a systematic inclusion of higher-order terms that are
omitted from standard-MCT-like theories \cite{das:12,das:13,mccowan:15}. It is
likely that such approaches can be formulated in a Martin-Schwinger-type form,
which would enable the direct comparison of the hierarchy presented here with
recent field-theoretic formulations.  Such a comparison might also make clearer
the effect of several approximations that have been made here. Future work
should be devoted to such a program.

\acknowledgments
We are greatly indebted to the authors of Ref.\ \cite{weysser:10}, in
particular Antonio M.\ Puertas and Matthias Fuchs, for generously making
available the results of their simulations to us.  LMCJ gratefully acknowledges
support from the Netherlands Organization for Scientific Research (NWO) through
a Rubicon fellowship.  DRR ackowledges grant NSF-CHE 1464802 for support.


\end{document}